MAGNETISM, MAGNETIC MATERIALS AND SPINTRONICS

# HIGH ISOTHERMAL MAGNETOCALORIC EFFECT IN La(Fe,Si)$_{13}$ BASED ALLOYS

A.P. Kamantsev[a,b], Yu. S. Koshkid'ko[a], O.E. Kovalev[b], N.Yu. Nyrkov[b], A.V. Golovchan[b], A.A. Amirov[c], A.M. Aliev[c]

[a] *Kotelnikov Institute of Radioengineering and Electronics of RAS, Moscow, 125009 Russia*
[b] *Galkin Donetsk Institute for Physics and Technology, Donetsk, 283048 Russia*
[c] *Amirkhanov Institute of Physics, Dagestan Federal Research Center of RAS, Makhachkala, Republic of Dagestan, 367003 Russia*
\* e-mail: kaman4@gmail.com

This work investigates the magnetocaloric effect (MCE) in LaFe$_{11.6}$Si$_{1.4}$ and LaFe$_{11.78}$Mn$_{0.41}$Si$_{1.32}$H$_{1.6}$ alloys under adiabatic ΔT and isothermal ΔQ conditions in a magnetic field of $\mu_0 H = 1.8$ T. The studied samples exhibited high reproducibility of the ΔQ-effect upon cyclic magnetic field application, which is of critical importance for magnetic cooling systems. The LaFe$_{11.78}$Mn$_{0.41}$Si$_{1.32}$H$_{1.6}$ alloy demonstrate high values of the isothermal MCE, with a maximum ΔQ = 3400 J/kg in a magnetic field of $\mu_0 H = 1.8$ T near the Curie temperature ($T_C$) of 275 K. This value is 2.5 times higher than the well-known corresponding values for pure Gd at room temperature. Furthermore, the structural and magnetic properties of LaFe$_{13-x}$Si$_x$-based alloys with Cr and Co additions were investigated using density functional theory calculations. It was shown that the addition of Cr leads to a decrease in the equilibrium volume, i.e., to a compression of the crystal lattice, whereas Co addition causes its expansion. These changes are expected to increase or decrease the $T_C$, respectively

## INTRODUCTION

The technology of magnetic cooling at room temperature based on solid-state magnetic materials with phase transitions (PT) is currently developing rapidly [1, 2], and the first commercial proposals for magnetic refrigerators based on pure Gd have appeared [3]. Nevertheless, a search is still underway for more energy-efficient and cheaper MCE materials than Gd, such as, for example, La-Fe-Si alloys [4]. Intermetallic alloys LaFe$_{13-x}$Si$_x$ are widely studied due to the significant MCE discovered in them, which is observed near room temperature in the concentration range of $1.0 \leqslant x \leqslant 1.8$ [5]. The advantage of this family of alloys is the ability to vary the Curie temperature $T_C$ in a wide range by both varying x and alloying with Ce, Pr, Nd [6], Cr [7], and Mn atoms, both in homogeneous samples [8-9] and in composites [10-11]. Hydrogenated compounds La(Fe,Mn,Si)$_{13}$H$_y$ exhibit $T_C$ in the room temperature region and show high values of adiabatic MCE near this PT [8-11].

Direct measurements of the MCE under adiabatic conditions in the LaFe$_{11.6}$Si$_{1.4}$ alloy considered in this work were carried out in a magnetic field of $\mu_0 H = 1.9$ T in [12], with the maximum MCE value being ΔT = 7 K at $T_0 = 194$ K in the sample cooling mode (Fig. 1). The values of the adiabatic MCE in this alloy are quite high and significantly (by 2.5 K) exceed similar MCE values for polycrystalline Gd in the same field at room temperature [13, 14].

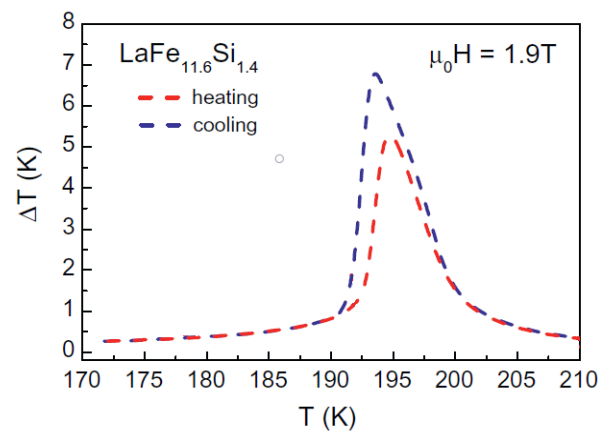

Fig. 1. Temperature dependences of the adiabatic temperature change ΔT for LaFe$_{11.6}$Si$_{1.4}$ annealed at 1323 K for 7 days, data from [12].

In this work, in addition to the LaFe$_{11.6}$Si$_{1.4}$ alloy samples, we also investigated samples of hydrogenated LaFe$_{11.78}$Mn$_{0.41}$Si$_{1.32}$H$_{1.6}$ alloys, both series produced by *Vacuumschmelze GmbH*, Germany. Direct measurements of the adiabatic MCE in the LaFe$_{11.78}$Mn$_{0.41}$Si$_{1.32}$H$_{1.6}$ alloy were carried out using a well-known technique in fields of $\mu_0 H = 1.8$ T using equipment from AMT&C LLC [15].

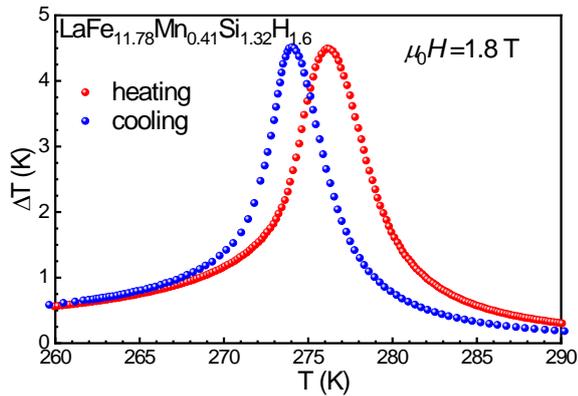

Fig. 2. Temperature dependences of the adiabatic temperature change ΔT for $LaFe_{11.78}Mn_{0.41}Si_{1.32}H_{1.6}$ during heating and cooling.

An important functional characteristic of magnetocaloric materials is the isothermal MCE [16], therefore, in this work, direct measurements of the isothermal heat ΔQ in a magnetic field were carried out in the described samples. A magnetic system from AMT&C LLC and a specially designed insert into it were used for the studies. The method for measuring ΔQ consists in the fact that the measurement of the temperature change $\Delta T_b$ is carried out not on the sample itself, but on a massive copper block with a known heat capacity $C_b$, independent of the magnetic field, while the block is brought into good thermal contact with the sample [17]. We can neglect the heat associated with the change in temperature of the sample if its mass m is negligibly small compared to the mass of the block $M_b \gg m$, then

$$\Delta Q \approx -\frac{M_b}{m} \cdot C_b \cdot \Delta T_b \quad (1)$$

The aim of this work was to directly investigate the isothermal MCE and its repeatability in $LaFe_{11.6}Si_{1.4}$ and $LaFe_{11.78}Mn_{0.41}Si_{1.32}H_{1.6}$ alloy samples in readily accessible magnetic fields of 1.8 T to demonstrate the applicability of the alloys in magnetic refrigeration technology. The second aim of this work was to theoretically investigate the properties of $LaFe_{13-x}Si_x$-based alloys, initiated in [11], aimed at further improving the functional characteristics of the alloys.

EXPERIMENTAL STUDIES

The results of direct measurements of the MCE in the $LaFe_{11.78}Mn_{0.41}Si_{1.32}H_{1.6}$ alloy under adiabatic conditions ΔT in a magnetic field of $\mu_0 H = 1.8$ T are shown in Fig. 2, with the maximum MCE value being ΔT = 4.5 K near TC = 275 K in the sample heating and cooling modes. The obtained ΔT values are lower than those of the $LaFe_{11.6}Si_{1.4}$ alloy (Fig. 1), since the magnetization of the ferromagnetic phase of $LaFe_{11.6}Si_{1.4}$ is higher due to a lower (by ~80 K) TC. The observed increase in $T_C$ during hydrogenation is associated with the growth of exchange interactions between the Fe atoms (Table 1) located at the vertices of the icosahedron [11]. However, the ΔT values for $LaFe_{11.78}Mn_{0.41}Si_{1.32}H_{1.6}$ are comparable with the values for Gd obtained in the same field at initial temperatures close to room temperature [13, 14].

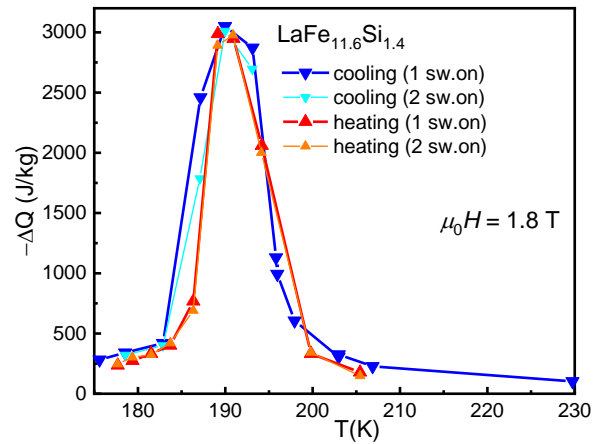

Fig. 3. Temperature dependences of isothermal heat release ΔQ of the $LaFe_{11.6}Si_{1.4}$ sample during cooling and heating. Measurements for each temperature were carried out twice (first and second field switching).

The results of measurements of the isothermal ΔQ-effect during sequential heating and cooling in the $LaFe_{11.6}Si_{1.4}$ sample are shown in Fig. 3. The ΔQ maxima near $T_C = 190$ K in different modes turned out to be approximately equal and amounted to ΔQ = 3000 J/kg in a field of $\mu_0 H = 1.8$ T. In addition, the effect of the second switching on of the magnetic field at the same temperature was studied in the $LaFe_{11.6}Si_{1.4}$ alloy sample. It can be seen from Fig. 3 that ΔQ changed insignificantly during the second switching on of the field, that is, the sample demonstrates high repeatability of this effect. Previously, cycling was carried out only for the adiabatic ΔT-effect in this alloy [18], and a significant decrease in the ΔT-effect in the sample cooling mode was shown.

The results of measurements of the isothermal ΔQ effect in the cooling and heating modes in the $LaFe_{11.78}Mn_{0.41}Si_{1.32}H_{1.6}$ sample are shown in Fig. 4. The ΔQ maxima near $T_C = 275$ K in different modes also were practically the same and amounted to ΔQ = 3400 J/kg in a magnetic field of $\mu_0 H = 1.8$ T. This result is quite consistent with the estimate given in [4] using an indirect method. The obtained ΔQ values are high, since they also exceed those of $LaFe_{11.6}Si_{1.4}$ (Fig. 3). Moreover, with comparable values of the ΔT-effect for Gd [13, 14] and the LaFe$_{11.78}$Mn$_{0.41}$Si$_{1.32}$H$_{1.6}$ alloy (Fig. 2), the values of the ΔQ-effect in the latter are 2.5 times higher (Fig. 4) in a magnetic field of up to 2 T, which is explained by the presence of latent heat of the magnetically induced first-order PT. Comparing the ΔQ values for LaFe$_{11.78}$Mn$_{0.41}$Si$_{1.32}$H$_{1.6}$ with the known values for Heusler alloys of the Ni-Mn-Z (Z=Ga, In) system [19-20], we can also conclude that alloys of the LaFe$_{13-x}$Si$_x$ system are much more effective in fields of up to 2 T.

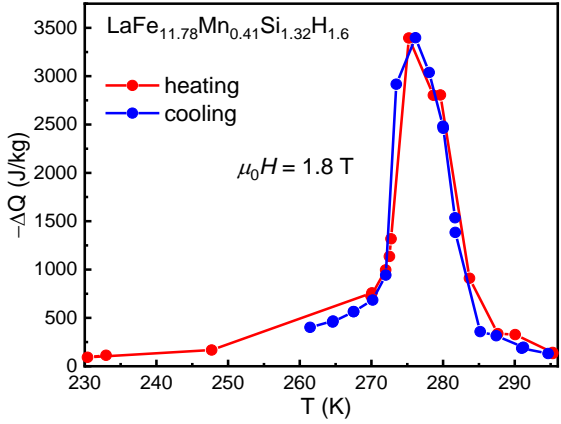

Fig. 4. Temperature dependences of isothermal heat release ΔQ of the LaFe$_{11.78}$Mn$_{0.41}$Si$_{1.32}$H$_{1.6}$ sample during cooling and heating.

Of interest is the virtual absence of hysteresis in the temperature dependences of ΔQ (Fig. 3 and Fig. 4), in contrast to the data for ΔT during heating and cooling (Fig. 1 and Fig. 2). For ΔT, the shift in the temperature of the maximum effect depending on the applied external field is observed in the vicinity of the first-order PT and is described by the Clausius-Clapeyron equation [17, 19]. Direct studies of ΔQ in the same materials (MnAs, Ni$_{2.18}$Mn$_{0.82}$Ga) showed a minimal discrepancy between the temperatures of the maximum effect during heating and cooling, as well as their correspondence to the temperature of the peak energy absorption on the differential scanning calorimetry curve during heating [17, 19]. This circumstance is explained by the fact that in all these alloys, the inclusion of a magnetic field causes a low-energy ferromagnetic phase, and therefore the peak of energy release in the isothermal regime will correspond to the peak of absorption during heating without a field.

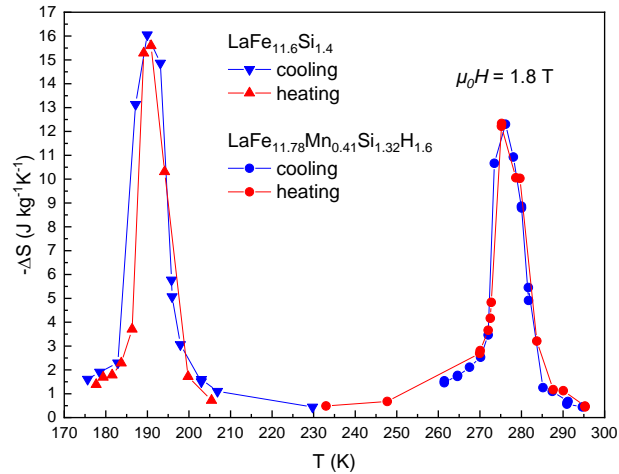

Fig. 5. Temperature dependences of the isothermal change in entropy ΔS for LaFe$_{11.6}$Si$_{1.4}$ and LaFe$_{11.78}$Mn$_{0.41}$Si$_{1.32}$H$_{1.6}$ during cooling and heating in a magnetic field of 1.8 T.

Since the change in magnetic entropy ΔS and the isothermal MCE are related by a simple relation

$$\Delta S = \frac{\Delta Q}{T}, \quad (2)$$

where T is the sample temperature, this material parameter can be easily estimated from direct experiments. The results of such an assessment for LaFe$_{11.6}$Si$_{1.4}$ and LaFe$_{11.78}$Mn$_{0.41}$Si$_{1.32}$H$_{1.6}$ alloy samples are shown in Figure 5.

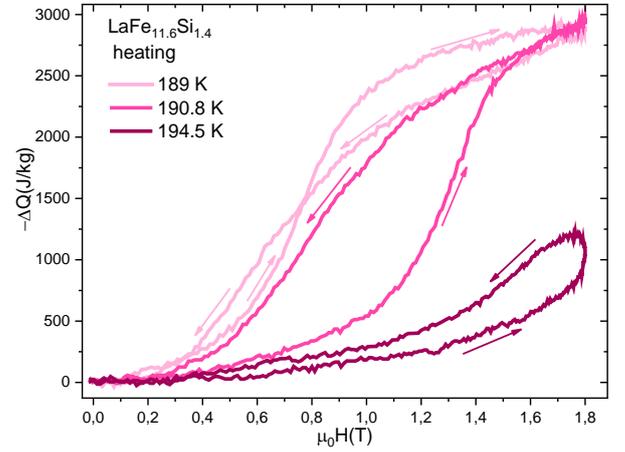

(a)

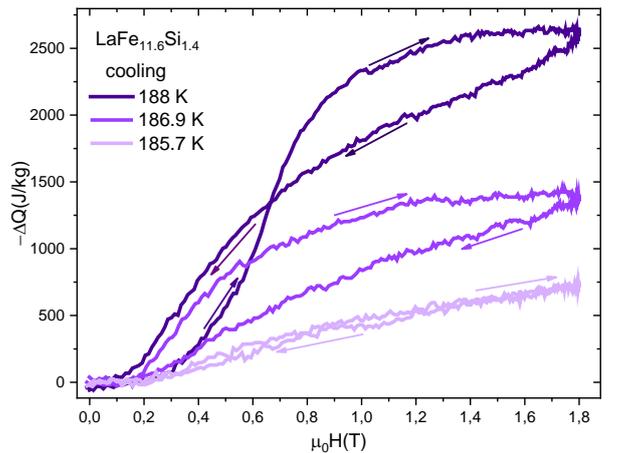

(b)

Fig. 6. Magnetic field dependences of the isothermal change in heat ΔQ for LaFe$_{11.6}$Si$_{1.4}$ at the first turn on of the field in the (a) heating and (b) cooling modes.

If we compare the obtained results with those known in the literature, then for LaFe$_{11.6}$Si$_{1.4}$ the maximum value is known to be ΔS = 21.7 J/(kg K) in a magnetic field of μ$_0$H = 2 T at 190 K [4, 18], and for La-Fe-Mn-Si-H alloys ΔS = 13 J/(kg K) is known in a magnetic field of μ$_0$H = 2 T at room temperature [4]. In general, for the LaFe$_{11.78}$Mn$_{0.41}$Si$_{1.32}$H$_{1.6}$ alloy we obtained similar results, and for LaFe$_{11.6}$Si$_{1.4}$ they are slightly lower than known.

It is also interesting to examine the curves obtained by directly measuring the isothermal heat release ΔQ as a function of the magnetic field for LaFe$_{11.6}$Si$_{1.4}$ (Fig. 6). Figure 6a shows the field dependences of ΔQ at different temperatures in the sample heating mode, and Fig. 6b – in the cooling mode. It is evident that in the heating mode (Fig. 6a), at 189-191 K in a field of 1.8 T, a magnetically induced first-order PT occurs, which causes high ΔQ values in the alloy. However, with a further increase in temperature to 194.5 K, the MCE occurs only due to the paraprocess in the paramagnetic phase, and the ΔQ value drops by a factor of 3. Similar behavior is also observed during measurements in the sample cooling mode (Fig. 6b) – at 188 K, high values of ΔQ are obtained due to the magnetically induced first-order PT, but with further cooling, the value of ΔQ drops significantly, since at 186 K magnetization of the ferromagnetic phase occurs.

## ELECTRONIC STRUCTURE AND INTERATOMIC EXCHANGE INTERACTIONS

In the present work, the effect of Cr and Co alloying on the electronic structure and parameters of interatomic exchange interactions was investigated using the fully relativistic SPRKKR v8.6 package [21, 22] within the framework of density functional theory. Preliminary calculations showed that the use of the GGA approximation for the exchange-correlation energy (in comparison with the LDA approximation) leads to an excessive overestimation of the "theoretical" Curie temperature; therefore, the exchange-correlation energy was subsequently calculated in the LDA approximation [23]. The atomic sphere approximation was used for the crystal potential. The electronic structure of the alloy was calculated in the coherent potential approximation for the disordered alloy model. LaFe$_{13-x}$Si$_x$ alloys have a cubic crystal structure of the NaZn$_{13}$ type (space group $Fm\overline{3}c$, Fig. 7) [24], in which Fe atoms occupy two types of positions Fe$_I$ – 8b(0,0,0) and Fe$_{II}$ – 96i(0,y,z). 12 Fe$_{II}$ atoms are located at the vertices of a regular icosahedron, in the center of which Fe$_I$ is located. It is assumed that Si atoms are uniformly distributed over the Fe$_{II}$ positions [24]. La atoms occupy positions of the 8a(1/4,1/4,1/4) type, the crystal lattice parameters are

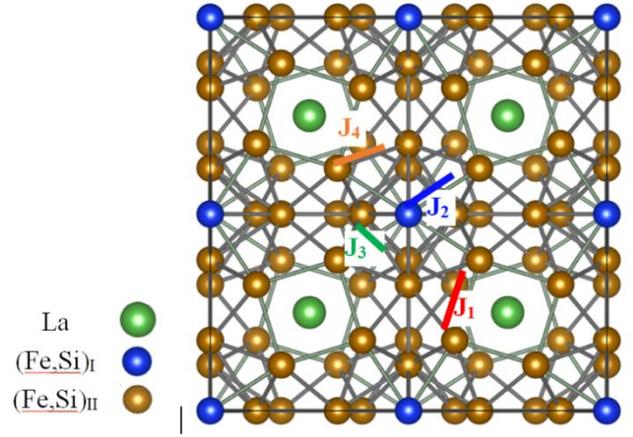

Fig. 7. Crystal structure of LaFe$_{13-x}$Si$_x$ and the main interatomic exchange integrals, where J$_1$ is the exchange integral between Fe$_{II}$-Fe$_{II}$ from neighboring icosahedrons, J$_2$ is the exchange integral between Fe$_I$-Fe$_{II}$ (between the shell of an icosahedron and its center), J$_3$ is the exchange integral between Fe$_{II}$-Fe$_{II}$ from neighboring icosahedra, J$_4$ is the exchange integral between Fe$_{II}$-Fe$_{II}$.

taken from [24] (a = 11.461 Å, y = 0.179, z = 0.1168 Å).

Interatomic exchange integrals were calculated using the technique [25, 26], based on the calculation of the second derivative of the total energy functional with respect to the deviations of a selected pair of spins from the equilibrium position. The FM spin configuration was chosen as the basis for calculating the exchange integrals. The technique used also allows one to calculate the parameters of the Dzyaloshinsky-Moriya interaction [27]. Thus, the system of itinerant electrons is replaced by an effective spin Hamiltonian of the form:

$$H = -\frac{1}{2}\sum_{i \neq j}\left(J_{ij}^{\alpha\beta}e_i^{\alpha}e_j^{\beta} + \vec{D}_{ij}\vec{e}_i \times \vec{e}_j\right), \quad (3)$$

Where $\vec{e}_i$ - is the unit vector indicating the direction of the magnetic moment at the site $i$, $J_{ij}^{\alpha\beta}$ are the coefficients of the exchange interaction matrix, $\alpha, \beta = x, y$ and $\vec{D}_{ij}$ are the parameters describing the Dzyaloshinskii-Moriya interaction. In the system under study, the $\vec{D}_{ij}$ and off-diagonal terms $J_{ij}^{xy}, J_{ij}^{yx}$ do not exceed 0.3 meV and are therefore not considered in further calculations.

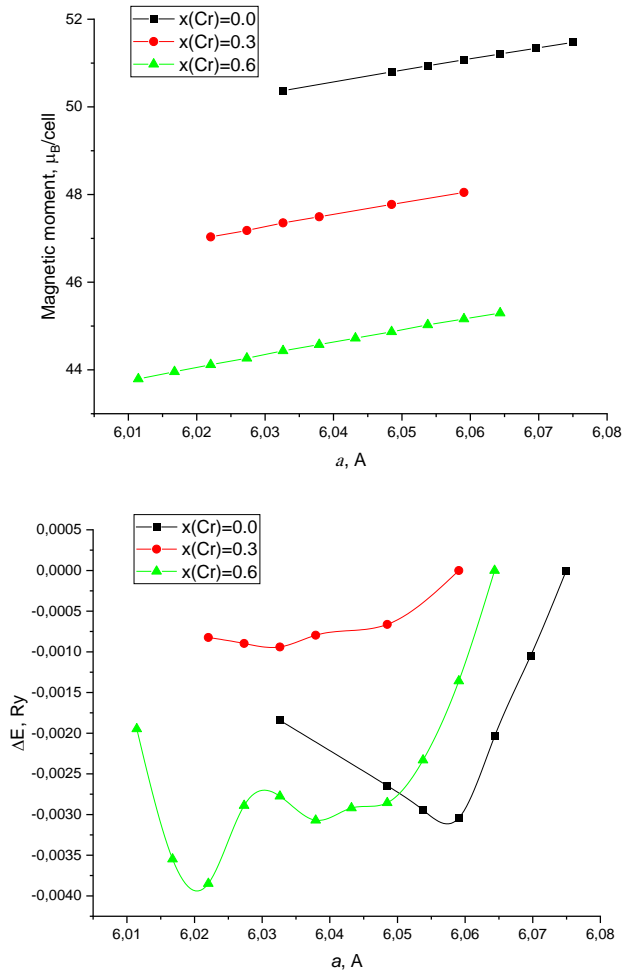

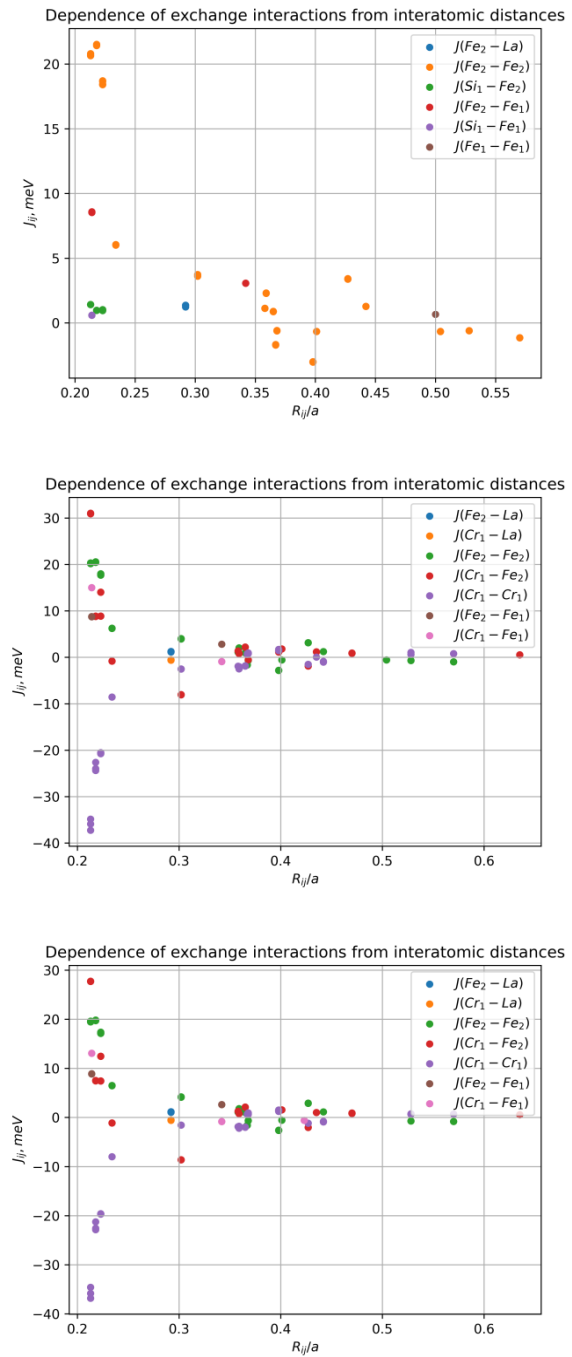

Fig. 8. The influence of Cr on the magnetic moment of the unit cell and the dependence of the total energy on the lattice constant in LaFe$_{11.596-y}$Cr$_y$Si$_{1.404}$.

According to the calculated dependence of the total energy of LaFe$_{11.596-y}$Cr$_y$Si$_{1.404}$ on the unit cell constant (Fig. 8), the addition of chromium leads to a decrease in the equilibrium volume, i.e., to a compression of the crystal lattice. In the ferromagnetic phase of LaFe$_{11.596-y}$Cr$_y$Si$_{1.404}$, the magnetic moment of a unit cell containing 28 atoms varies from 51.5 $\mu_B$ for LaFe$_{11.596}$Si$_{1.404}$ to 43.8 $\mu_B$ for LaFe$_{10.996}$Cr$_{0.6}$Si$_{1.404}$ (Fig. 8). The magnetic moments of the atoms vary from -0.4 $\mu_B$ to -0.45 $\mu_B$ for La, from 1.87 $\mu_B$ to 2.03 $\mu_B$ for Fe in the Fe$_I$ position, from 2.16 $\mu_B$ to 2.3 $\mu_B$ for Fe in the Fe$_{II}$ position, from -1.68 $\mu_B$ to -1.85 $\mu_B$ for Cr in the Fe$_{II}$ position, and from -0.158 $\mu_B$ to -0.19$\mu_B$ for Si, which is consistent with both the FPLAPW calculation data [28] for LaFe$_{11.31}$Si$_{1.69}$ (M(Fe$_I$) = 2.07 $\mu_B$, M(Fe$_{II}$) = 2.42 $\mu_B$) and the neutron diffraction results [29] for LaFe$_{11.4}$Si$_{1.6}$ (M(Fe$_I$) = 1.59 $\mu_B$, M(Fe$_{II}$) = 2.12 $\mu_B$).

Typical dependences of the main exchange integrals on the interatomic distance for LaFe$_{11.596-y}$Cr$_y$Si$_{1.404}$ are shown in Fig. 9. As can be seen from the figure, the exchange integrals decrease rather rapidly with increasing interatomic distance and do not exceed 1 meV already at a distance of 0.5 *a*. The most significant are the exchange interactions between Fe$_{II}$ atoms (~22 meV). The Fe$_{II}$-Fe$_I$ interaction (between the ico-sahedron shell and its center) is approximately 3 times weaker (~7 meV). The magnetic moments of the chromium atoms are directed opposite to the resulting magnetic moment of the unit cell. The magnitude of the exchange interaction between the nearest Cr and $Fe_{II}$ atoms is ~30 meV in $LaFe_{11.296}Cr_{0.3}Si_{1.404}$ and decreases insignificantly with lattice compression or an increase in the Cr concentration. The Cr-Cr exchange interaction is also quite large (from -20 to -37 meV) and increases with increasing chromium concentration. Thus, chromium doping in the $LaFe_{11.596-y}Cr_ySi_{1.404}$ system leads to a decrease in the total magnetization and promotes a decrease in $T_C$, which was also observed experimentally in [7].

Fig. 9. Dependence of the values of the interatomic interaction integrals on the interatomic distance for LaFe$_{11.596}$Si$_{1.404}$ (top), LaFe$_{11.296}$Cr$_{0.3}$Si$_{1.404}$ (middle) and LaFe$_{10.996}$Cr$_{0.6}$Si$_{1.404}$ (bottom). In all cases, the crystal lattice constant a = 6.06 Å.

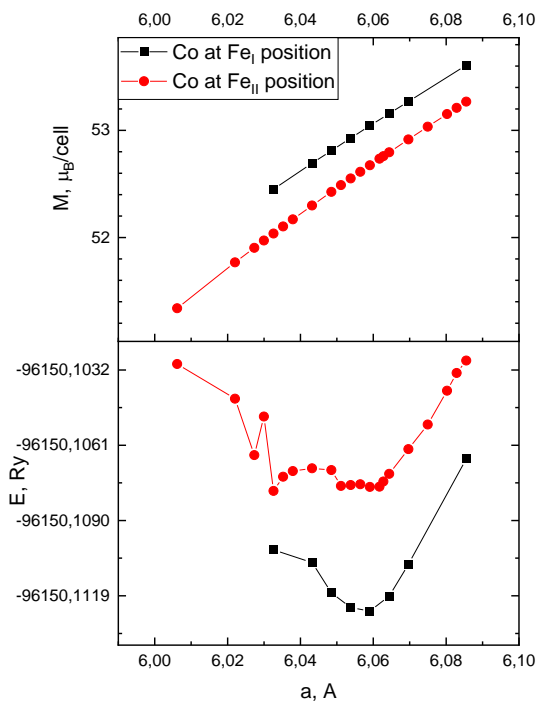

Fig. 10. The influence of the distribution of Co over iron positions on the dependence of the magnetic moment of the unit cell and the total energy on the lattice constant.

Unlike Cr, which has two 3d electrons less than iron, Co has one more 3d electron and, accordingly, a larger ionic radius, which means that the addition of Co should lead to an increase in the unit cell size and an increase in $T_C$, which was observed experimentally in [32]. We calculated the dependence of the total energy on the unit cell size for two variants of the Co atom arrangement. In the first, they were considered randomly distributed over the $Fe_I$ positions, and in the second, it was assumed that the Co atoms were randomly distributed over the $Fe_{II}$ positions. A compari-son of the calculation results for $LaFe_{11.248}Co_{0.648}Si_{1.104}$ is shown in Fig. 10. As can be seen from the figure, it is energetically favorable for Co to occupy $Fe_I$-type positions. At the equilibrium volume in the ferromagnetic phase $LaFe_{11.248}Co_{0.648}Si_{1.104}$ the magnetic moment of the unit cell is 53.04 $\mu_B$, the magnetic moment of La is -0.485 $\mu_B$, $Fe_I$ 2.0 $\mu_B$, 2.34 $\mu_B$ for $Fe_{II}$, -0.196 $\mu_B$ for Si. The magnetic moment of Co atoms is co-directed with the magnetic moment of the unit cell and is 1.467 $\mu_B$.

Typical dependences of the main exchange integrals on the interatomic distance for $LaFe_{11.248}Co_{0.648}Si_{1.104}$ are shown in Fig. 11. As can be seen from the figure, the exchange integrals decrease quite rapidly with increasing interatomic distance and do not exceed 1 meV already at a distance of 0.5 $a$. The most significant exchange interactions are between $Fe_{II}$ atoms (~25 meV). The $Co_I$-$Fe_{II}$ exchange interaction lags slightly behind (~22 meV). The $Fe_{II}$-$Fe_I$ interaction (between the icosahedron shell and its center) is approximately 3 times weaker (~8 meV). The Co-Co exchange interaction is also quite strong (~22 meV), and changes in the unit cell volume have little effect

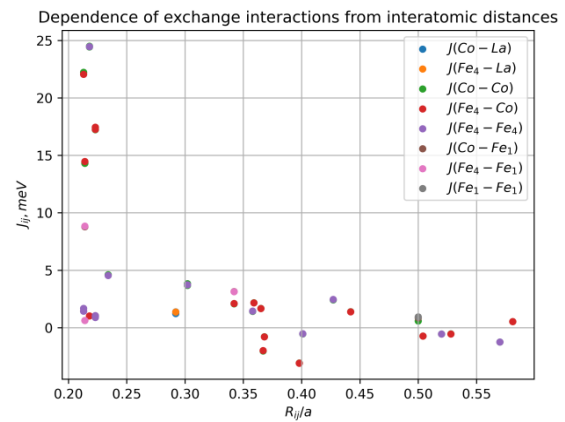

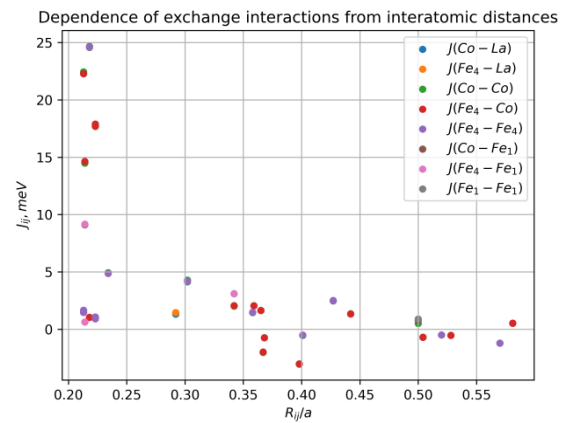

Fig. 11. Dependence of the values of the integrals of interatomic interactions on the interatomic distance for $La(Fe_{0.352}Co_{0.648})_1(Fe_{0.908}Si_{0.092})_{12}$ with a lattice constant a=6.033 Å (top) and a=6.059 Å (bottom).

on interatomic exchange interactions (Fig. 11).

## CONCLUSION

Samples of LaFe$_{11.6}$Si$_{1.4}$ and LaFe$_{11.78}$Mn$_{0.41}$Si$_{1.32}$H$_{1.6}$ alloys demonstrated high repeatability of the ΔQ effect upon repeated magnetic field application, which is critical for magnetic refrigeration systems. LaFe$_{11.78}$Mn$_{0.41}$Si$_{1.32}$H$_{1.6}$ samples exhibit high isothermal MCE values with a maximum ΔQ of 3400 J/kg in a magnetic field of $\mu_0 H$ = 1.8 T near TC = 275 K, which is 2.5 times higher than similar known values for pure Gd at room temperature.

To study the possibilities of further improving the functional properties of LaFe$_{13-x}$Si$_x$-based alloys, calculations were performed for the addition of Co and Cr. Based on the calculations, it was shown that the addition of Cr leads to a decrease in the equilibrium volume, i.e., to a compression of the crystal lattice, which is consistent with both the FPLAPW calculation data and the neutron diffraction results. Unlike Cr, which has two 3d electrons less than iron, Co has one more 3d electron and, accordingly, a larger ionic radius, which means that the addition of Co should lead to an increase in the unit cell size. Thus, the addition of Cr leads to a decrease, and Co to an increase in TC in these alloys, which is confirmed experimentally in [7] and [32], respectively.

## ACKNOWLEDGEMENTS

The authors thank K.P. Skokov for providing the samples for the study.

There are no conflicts of interest.

МАГНЕТИЗМ, МАГНИТНЫЕ МАТЕРИАЛЫ И СПИНТРОНИКА

УДК

# ВЫСОКИЙ ИЗОТЕРМИЧЕСКИЙ МАГНИТОКАЛОРИЧЕСКИЙ ЭФФЕКТ В СПЛАВАХ НА ОСНОВЕ La(Fe,Si)$_{13}$

А.П. Каманцев[a,b]*, Ю.С. Кошкидько[a], О.Е. Ковалёв[b], Н.Ю. Нырков[b], А.В. Головчан[b], А.А. Амиров[c], А.М. Алиев[c]

[a] *Институт радиотехники и электроники им. В.А.Котельникова РАН, ул. Моховая, 11, корп. 7, Москва, 125009 Россия*

[b] *ФГБНУ «Донецкий физико-технический институт им. А.А. Галкина», ул. Р. Люксембург, 72, Донецк, 283048 Россия*

[c] *Институт физики им. Х.И. Амирханова ДФИЦ РАН, ул. М. Ярагского, 94, Махачкала, 367003 Россия*

* e-mail: kaman4@gmail.com

В работе исследован магнитокалорический эффект (МКЭ) сплавов LaFe$_{11.6}$Si$_{1.4}$ и LaFe$_{11.78}$Mn$_{0.41}$Si$_{1.32}$H$_{1.6}$ в адиабатических $\Delta T$ и изотермических $\Delta Q$ условиях в магнитном поле $\mu_0 H = 1.8$ Тл. Исследуемые образцы показали высокую повторяемость $\Delta Q$-эффекта при повторном включении магнитного поля, что критически важно для систем магнитного охлаждения. Сплав LaFe$_{11.78}$Mn$_{0.41}$Si$_{1.32}$H$_{1.6}$ демонстрирует высокие значения изотермического МКЭ с максимальным $\Delta Q = 3400$ Дж/кг в магнитном поле $\mu_0 H = 1.8$ Тл вблизи $T_C = 275$ К, что в 2.5 раза превосходит аналогичные известные значения для чистого Gd при комнатной температуре. Также выполнены исследования структурных и магнитных свойств сплавов на основе LaFe$_{13-x}$Si$_x$ с добавлением Cr и Co методами теории функционала электронной плотности. Показано, что добавление Cr приводит к уменьшению равновесного объема, т.е. к сжатию кристаллической решетки, а Co – к её расширению, что должно приводить к росту или снижению $T_C$, соответственно.

*Ключевые слова:* магнитокалорический эффект, сплавы La-Fe-Si, сильные магнитные поля, изотермическое тепло

## ВВЕДЕНИЕ

Технология магнитного охлаждения при комнатной температуре на основе твердотельных магнитных материалов с фазовыми переходами (ФП) в настоящее время стремительно развивается [1, 2], появились первые коммерческие предложения магнитных холодильников, созданных на основе чистого Gd [3]. Тем не менее, до сих пор ведутся поиски более энергоэффективных и дешевых чем Gd материалов с МКЭ, какими, например, являются сплавы системы La-Fe-Si [4]. Интерметаллические сплавы LaFe$_{13-x}$Si$_x$ широко исследуются из-за обнаруженного в них значительного МКЭ, который в интервале концентраций $1.0 \leqslant x \leqslant 1.8$ наблюдается вблизи комнатной температуры [5]. Достоинством данного семейства сплавов является возможность изменять температуру Кюри $T_C$ в широком диапазоне как изменяя $x$, так и легируя атомами Ce, Pr, Nd [6], Cr [7], а также Mn, как в однородных образцах [8-9], так и в композитах [10-11]. Гидрированные соединения La(Fe,Mn,Si)$_{13}$H$_y$ проявляют $T_C$ в области комнатных температур и показывают высокие значения адиабатического МКЭ вблизи этого ФП [8-11].

Прямые измерения МКЭ в адиабатических условиях в рассматриваемом в данной работе сплаве LaFe$_{11.6}$Si$_{1.4}$ проводились в магнитном поле $\mu_0 H = 1.9$ Тл в работе [12], при этом максимальное значение МКЭ составило $\Delta T = 7$ К при $T_0 = 194$ К в режиме охлаждения образца (Рис. 1). Величины адиабатического МКЭ в данном сплаве являются довольно высокими, и значительно (на 2.5 К) превышают аналогичные значения МКЭ для поликристаллического Gd в том же поле при комнатной температуре [13, 14].

В настоящей работе кроме образцов сплава LaFe$_{11.6}$Si$_{1.4}$ исследовалась образцы гидрированных сплавов LaFe$_{11.78}$Mn$_{0.41}$Si$_{1.32}$H$_{1.6}$, обе серии производства *Vacuumschmelze GmbH*, Германия. Прямые измерения адиабатического МКЭ в сплаве LaFe$_{11.78}$Mn$_{0.41}$Si$_{1.32}$H$_{1.6}$ проводились по известной методике в полях $\mu_0 H = 1.8$ Тл на оборудовании компании LLC "AMT&C" [15].

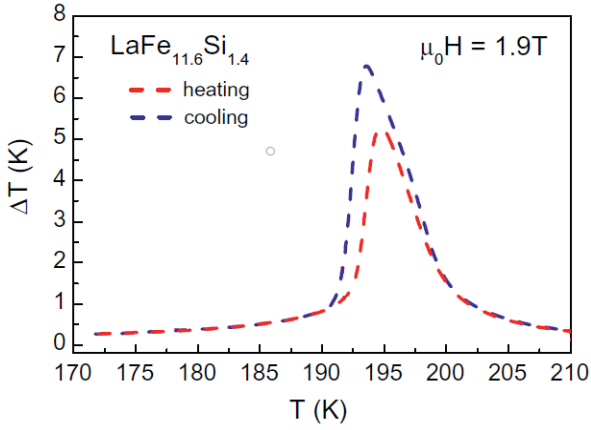

Рис. 1. Температурные зависимости адиабатического изменения температуры $\Delta T$ для $LaFe_{11.6}Si_{1.4}$, отожженного при 1323 K в течение 7 суток, данные из [12].

Важной функциональной характеристикой магнитокалорических материалов является изотермический МКЭ [16], поэтому в данной работе проводились прямые измерения изотермического тепла $\Delta Q$ в магнитном поле в описанных образцах. Для исследований использовалась магнитная система от LLC "AMT&C" и специально созданная вставка в нее. Метод измерения $\Delta Q$ заключается в том, что измерение изменения температуры $\Delta T_b$ ведётся не на самом образце, а на массивном медном блоке с известной теплоемкостью $C_b$, не зависящей от магнитного поля, при этом блок приводится в хороший тепловой контакт с образцом [17]. Мы можем пренебречь теплотой, связанной с изменением температуры образца, если его масса $m$ будет пренебрежимо мала по сравнению с массой блока $M_b \gg m$, тогда

$$\Delta Q \approx -\frac{M_b}{m} \cdot C_b \cdot \Delta T_b \qquad (1)$$

Целью данной работы было прямое исследование изотермического МКЭ и его повторяемости в образцах сплавов $LaFe_{11.6}Si_{1.4}$ и $LaFe_{11.78}Mn_{0.41}Si_{1.32}H_{1.6}$ в легко доступных по величине магнитных полях в 1.8 Тл, чтобы показать применимость сплавов в технологии магнитного охлаждения. Второй целью данной работы было теоретическое исследование свойств сплавов на основе $LaFe_{13-x}Si_x$, начатое в работе [11], направленное на дальнейшее улучшение функциональных характеристик сплавов.

## ПРЯМЫЕ ЭКСПЕРИМЕНТАЛЬНЫЕ ИССЛЕДОВАНИЯ

Результаты прямых измерений МКЭ в сплаве $LaFe_{11.78}Mn_{0.41}Si_{1.32}H_{1.6}$ в адиабатических условиях $\Delta T$ в магнитном поле $\mu_0H = 1.8$ Тл представлены на Рис. 2, при этом максимальное значение МКЭ

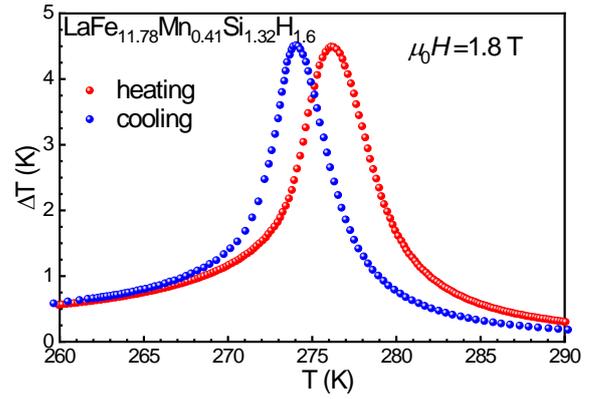

Рис. 2. Температурные зависимости адиабатического изменения температуры $\Delta T$ для $LaFe_{11.78}Mn_{0.41}Si_{1.32}H_{1.6}$ при нагреве и охлаждении.

составило $\Delta T = 4.5$ K вблизи $T_C = 275$ K в режимах нагревах и охлаждения образца. Полученные величины $\Delta T$ ниже, чем у сплава $LaFe_{11.6}Si_{1.4}$ (Рис. 1), поскольку намагниченность ферромагнитной фазы $LaFe_{11.6}Si_{1.4}$ выше, благодаря более низкой (на ~80 K) $T_C$. Наблюдаемый рост $T_C$ при гидрировании связан с ростом обменных взаимодействий между атомами Fe (Табл.1), расположенными в вершинах икосаэдра [11]. Однако значения $\Delta T$ для $LaFe_{11.78}Mn_{0.41}Si_{1.32}H_{1.6}$ сравнимы со значениями для Gd, полученным в том же поле при начальных температурах близких к комнатной [13, 14].

Результаты измерений изотермического $\Delta Q$-эффекта при последовательных нагреве и охлаждении в образце $LaFe_{11.6}Si_{1.4}$ представлены на Рис. 3. Максимумы $\Delta Q$ вблизи $T_C = 190$ K в разных режимах оказались примерно равны и составили $\Delta Q = 3000$ Дж/кг в поле $\mu_0H = 1.8$ Тл. Кроме того, в образце сплава $LaFe_{11.6}Si_{1.4}$ исследовался эффект второго включения магнитного поля при той же температуре, из Рис. 3 видно, что $\Delta Q$ при втором

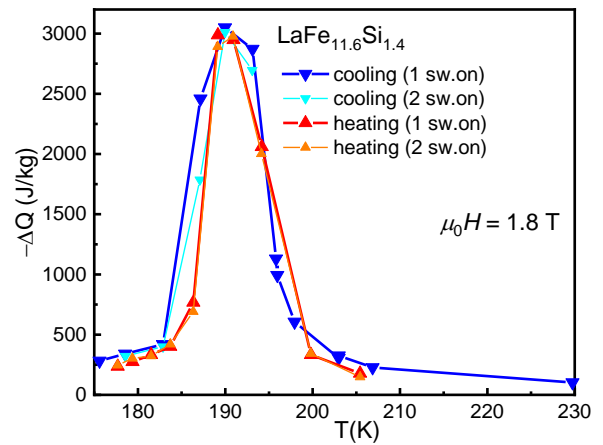

Рис. 3. Температурные зависимости изотермического тепловыделения $\Delta Q$ образца $LaFe_{11.6}Si_{1.4}$ при охлаждении и нагреве. Измерения для каждой температуры проводились дважды (первое и второе включение поля).

включении поля изменялся незначительно, то есть образец демонстрирует высокую повторяемость этого эффекта. Ранее проводилось циклирование только для адиабатического $\Delta T$-эффекта в данном сплаве [18] и было показано значительное снижение $\Delta T$-эффекта в режиме охлаждения образца.

Результаты измерений изотермического $\Delta Q$-эффекта в режимах охлаждения и нагрева в образце $LaFe_{11.78}Mn_{0.41}Si_{1.32}H_{1.6}$ представлены на Рис. 4. Максимумы $\Delta Q$ вблизи $T_C = 275$ К в разных режимах также практически не отличались и составили $\Delta Q = 3400$ Дж/кг в магнитном поле $\mu_0 H = 1.8$ Тл. Данный результат вполне соответствует оценке, данной в [4] косвенным методом. Полученные значения $\Delta Q$ являются высокими, так как превосходят и $LaFe_{11.6}Si_{1.4}$ (Рис 3). Кроме того, при сравнимых значениях $\Delta T$-эффекта для Gd [13, 14] и сплава $LaFe_{11.78}Mn_{0.41}Si_{1.32}H_{1.6}$ (Рис. 2), значения $\Delta Q$-эффекта в последнем в 2.5 раза выше (Рис. 4) в магнитном поле до 2 Тл, что объясняется наличием скрытой теплоты магнитоиндуцированного ФП 1-го рода. Сравнивая значения $\Delta Q$ для $LaFe_{11.78}Mn_{0.41}Si_{1.32}H_{1.6}$ с известными значениями для сплавов Гейслера системы Ni-Mn-Z (Z=Ga, In) [19-20], можно также сделать вывод, что сплавы системы $LaFe_{13-x}Si_x$ гораздо более эффективны в полях до 2 Тл.

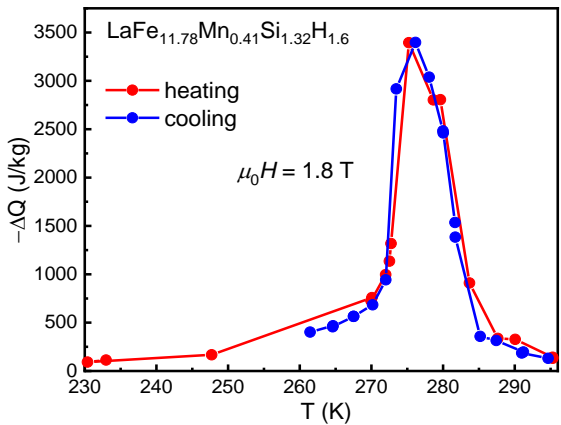

Рис. 4. Температурные зависимости изотермического тепловыделения $\Delta Q$ образца $LaFe_{11.78}Mn_{0.41}Si_{1.32}H_{1.6}$ при охлаждении и нагреве.

Вызывает интерес практическое отсутствие гистерезиса в зависимостях $\Delta Q$ от температуры (Рис. 3 и Рис. 4), в отличие от данных для $\Delta T$ при нагреве и охлаждении (Рис. 1 и Рис. 2). Для $\Delta T$ смещение температур максимума эффекта в зависимости от прикладываемого внешнего поля наблюдается в окрестности ФП 1-го рода и описывается уравнением Клапейрона-Клаузиуса [17, 19]. Прямые исследования $\Delta Q$ в этих же материалах (MnAs, $Ni_{2.18}Mn_{0.82}Ga$) показали минимальное расхождение температур максимума эффекта при нагреве и охлаждении, а также их соответствие температуре пика поглощения энергии на кривой дифференциальной сканирующей калориметрии при нагревании [17, 19]. Данное обстоятельство объясняется тем, что во всех этих сплавах включение магнитного поля вызывает низкоэнергетическую ферромагнитную фазу, и пик поэтому выделения энергии в изотермическом режиме будет соответствовать пику поглощения при нагревании без поля.

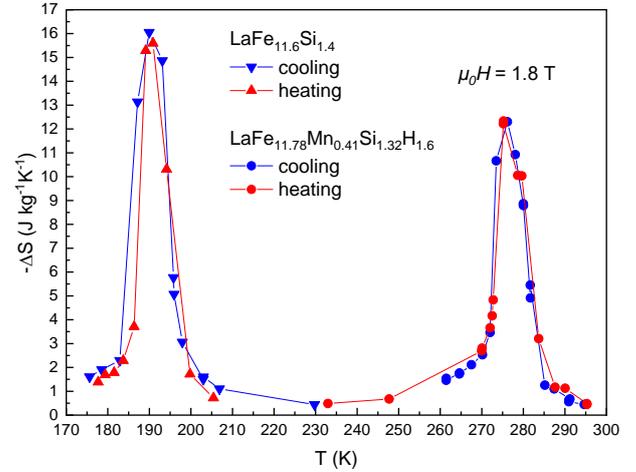

Рис. 5. Температурные зависимости изотермического изменения энтропии $\Delta S$ для $LaFe_{11.6}Si_{1.4}$ и $LaFe_{11.78}Mn_{0.41}Si_{1.32}H_{1.6}$ при охлаждении и нагреве в магнитном поле 1.8 Тл.

Поскольку изменение магнитной энтропии $\Delta S$ и изотермического МКЭ связаны простым соотношением

$$\Delta S = \frac{\Delta Q}{T}, \qquad (2)$$

где $T$ – температура образца, то из прямых экспериментов можно легко оценить и этот параметр материала. Результаты такой оценки для образцов сплавов $LaFe_{11.6}Si_{1.4}$ и $LaFe_{11.78}Mn_{0.41}Si_{1.32}H_{1.6}$ представлены на Рис. 5.

Если сравнивать полученные результаты с известными в литературе, то для $LaFe_{11.6}Si_{1.4}$ известно максимальное значение $\Delta S = 21.7$ Дж/(кг К) в магнитном поле $\mu_0 H = 2$ Тл при 190 К [4, 18], а для сплавов La-Fe-Mn-Si-H известно $\Delta S = 13$ Дж/(кг К) в магнитном поле $\mu_0 H = 2$ Тл при комнатной температуре [4]. В целом для сплава $LaFe_{11.78}Mn_{0.41}Si_{1.32}H_{1.6}$ мы получили схожие результаты, а для $LaFe_{11.6}Si_{1.4}$ несколько ниже известных.

Также интересно рассмотреть кривые, полученные при прямом измерении изотермического тепловыделения $\Delta Q$ в зависимости от магнитного поля для LaFe$_{11.6}$Si$_{1.4}$ (Рис. 6). На Рис. 6а представлены полевые зависимости $\Delta Q$ при разных температурах в режиме нагрева образца, а на Рис. 6b – в режиме охлаждения. Видно, что в режиме нагрева (Рис. 6а), при 189-191 К в поле 1.8 Тл происходит магнитоиндуцированный ФП 1-го рода, что и вы-

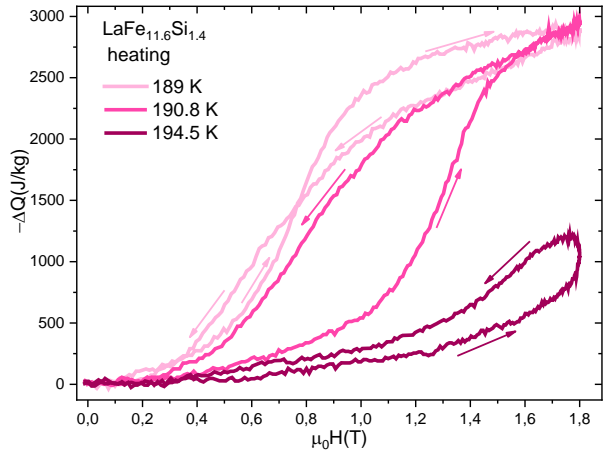

(a)

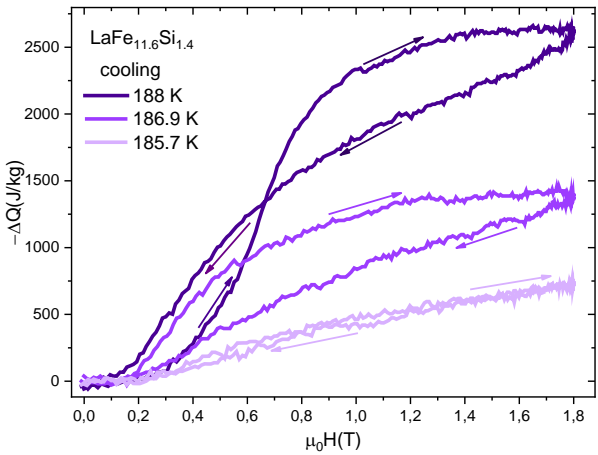

(b)

Рис. 6. Магнитополевые зависимости изотермического изменения тепла $\Delta Q$ для LaFe$_{11.6}$Si$_{1.4}$ при 1-ом включении поля в режимах (a) нагрева и (b) охлаждения.

зывает высокие значения $\Delta Q$ в сплаве, однако, при дальнейшем повышении температуры до 194.5 К МКЭ происходит только за счёт парапроцесса в парамагнитной фазе, и значение $\Delta Q$ падает в 3 раза. Аналогичное поведение наблюдается и при измерениях в режиме охлаждения образца (Рис. 6b) – при 188 К за счёт магнитоиндуцированного ФП 1-го рода получены высокие значения $\Delta Q$, но при дальнейшем охлаждении величина значительно падает $\Delta Q$, поскольку при 186 К идет намагничивание ферромагнитной фазы.

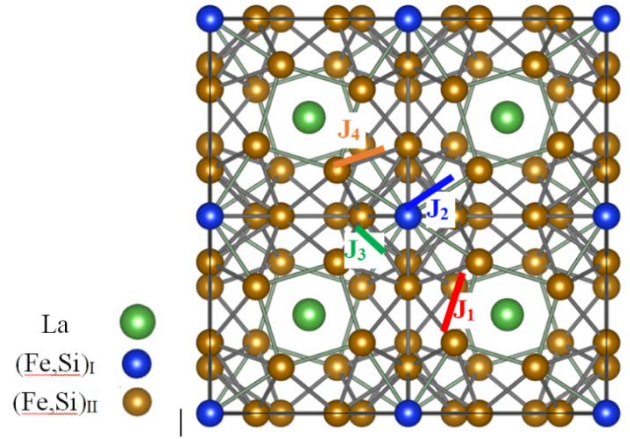

Рис. 7. Кристаллическая структура LaFe$_{13-x}$Si$_x$ и основные межатомные обменные интегралы, где $J_1$ – обменный интеграл между Fe$_{II}$-Fe$_{II}$ из соседних икосаэдров, $J_2$ – обменный интеграл между Fe$_I$-Fe$_{II}$ (между оболочкой икосаэдра и его центром), $J_3$ – это обменный интеграл между Fe$_{II}$-Fe$_{II}$ из соседних икосаэдров, $J_4$ – обменный интеграл между Fe$_{II}$-Fe$_{II}$.

## ЭЛЕКТРОННАЯ СТРУКТУРА И МЕЖАТОМНЫЕ ОБМЕННЫЕ ВЗАИМОДЕЙСТВИЯ

В представленной работе с помощью полнорелятивистского пакета SPRKKR v8.6 [21, 22] в рамках теории функционала электронной плотности исследовано влияние легирования Cr и Co на электронную структуру и параметры межатомных обменных взаимодействий. Предварительные расчеты показали, что использование GGA приближения для обменно-корреляционной энергии (в сравнении с LDA приближением) приводит к чрезмерному завышению «теоретической» температуры Кюри, поэтому в дальнейшем обменно-корреляционную энергию вычисляли в LDA приближении [23]. Для кристаллического потенциала использовали приближение атомных сфер. Расчет электронной структуры сплава проводили в приближении когерентного потенциала для модели неупорядоченного сплава. Сплавы LaFe$_{13-x}$Si$_x$ имеют кубическую кристаллическую структуру типа NaZn$_{13}$ (пространственная группа $Fm\overline{3}c$, рис. 7) [24], в которой атомы Fe занимают два типа позиций Fe$_I$ – $8b$(0,0,0) и Fe$_{II}$ – $96i$(0,y,z). 12 атомов Fe$_{II}$ находятся в вершинах правильного икосаэдра, в центре которого находится Fe$_I$. Предполагается, что атомы Si равномерно распределены по позициям Fe$_{II}$ [24]. Атомы La занимают позиции типа $8a$(1/4,1/4,1/4), параметры кристаллической решетки взяты из [24] ($a$ = 11.461 Å, y = 0.179, z = 0.1168 Å).

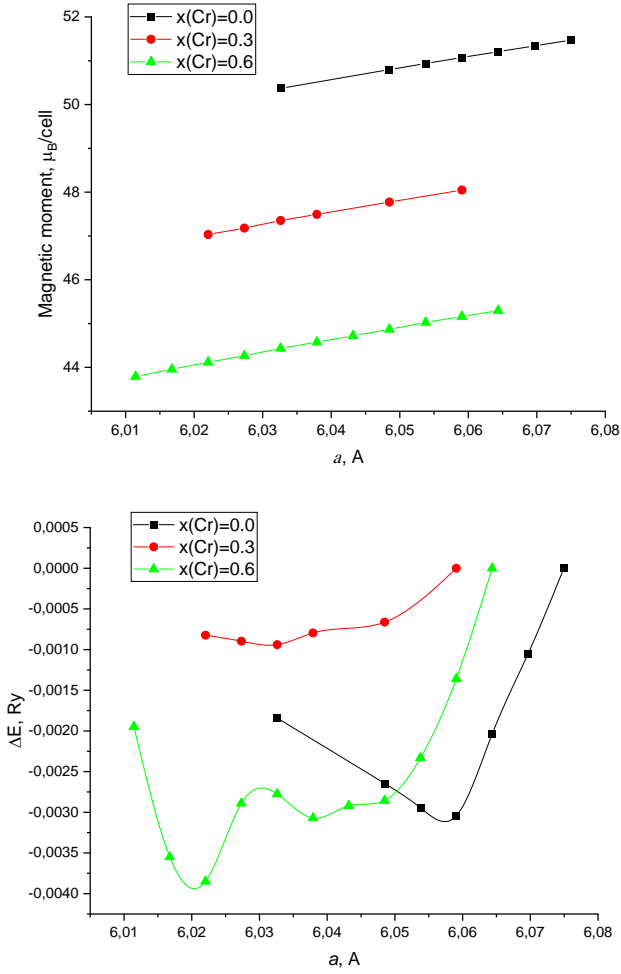

Рис. 8. Влияние Cr на магнитный момент элементарной ячейки и зависимость полной энергии от постоянной кристаллической решетки в LaFe$_{11.596-y}$Cr$_y$Si$_{1.404}$.

Межатомные обменные интегралы рассчитывали по методике [25, 26], основанной на расчете второй производной функционала полной энергии по отклонениям избранной пары спинов от положения равновесия. В качестве основной для расчета обменных интегралов выбрана ФМ конфигурация спинов. Используемая методика также позволяет рассчитывать параметры взаимодействия Дзялошинского-Мория [27]. Таким образом, система коллективизированных электронов заменяется эффективным спиновым гамильтонианом вида:

$$H = -\frac{1}{2}\sum_{i\neq j}\left(J_{ij}^{\alpha\beta}e_i^{\alpha}e_j^{\beta} + \vec{D}_{ij}\vec{e}_i\times\vec{e}_j\right), \quad (3)$$

где $\vec{e}_i$ – единичный вектор, указывающий направление магнитного момента на узле $i$, $J_{ij}^{\alpha\beta}$ –коэффициенты матрицы обменного взаимодействия, $\alpha, \beta = x, y$, $\vec{D}_{ij}$ – параметры описывающие взаимодействие Дзялошинского-Мория. В иссле-дуемой системе $\vec{D}_{ij}$ и недиагональные члены $J_{ij}^{xy}, J_{ij}^{yx}$ не превышают 0.3 мэВ, поэтому в дальнейших расчетах не учитываются.

Согласно результатам расчета зависимости полной энергии LaFe$_{11.596-y}$Cr$_y$Si$_{1.404}$ от постоянной элементарной ячейки (рис.8), добавление хрома приводит к уменьшению равновесного объема, т.е. к сжатию

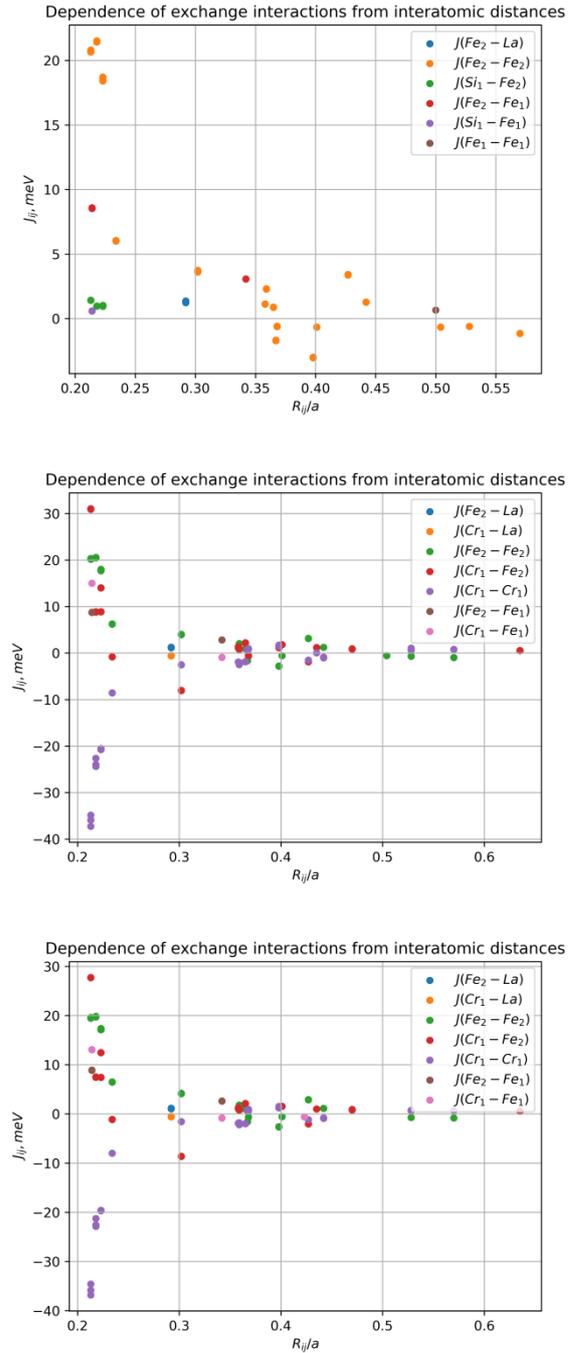

Рис.9. Зависимость величин интегралов межатомных взаимодействий от межатомного расстояния для LaFe$_{11.596}$Si$_{1.404}$ (сверху), LaFe$_{11.296}$Cr$_{0.3}$Si$_{1.404}$ (в середине) и LaFe$_{10.996}$Cr$_{0.6}$Si$_{1.404}$(снизу). Во всех случаях постоянная кристаллической решетки $a$ = 6.06 Å.

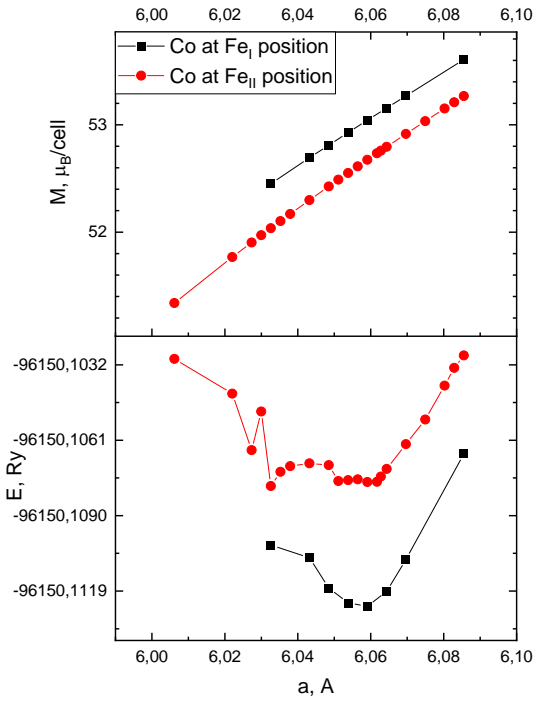

Рис. 10. Влияние распределения Co по позициям железа на зависимость магнитного момента элементарной ячейки и полной энергии от постоянной кристаллической решетки.

кристаллической решетки. В ферромагнитной фазе LaFe$_{11.596-y}$Cr$_y$Si$_{1.404}$ магнитный момент элементарной ячейки, содержащей 28 атомов, изменяется от 51.5 $\mu_B$ для LaFe$_{11.596}$Si$_{1.404}$ до 43.8 $\mu_B$ для LaFe$_{10.996}$Cr$_{0.6}$Si$_{1.404}$(рис.8). Магнитные моменты атомов изменяются от -0.4 $\mu_B$ до -0.45 $\mu_B$ для La, от 1.87 $\mu_B$ до 2.03 $\mu_B$ для Fe в позиции Fe$_I$, от 2.16 $\mu_B$ до 2.3 $\mu_B$ для Fe в позиции Fe$_{II}$, от -1.68 $\mu_B$ до -1.85 $\mu_B$ для Cr в позиции Fe$_{II}$, и от -0.158 $\mu_B$ до -0.19 $\mu_B$ для Si, что согласуется как с данными расчета методом FPLAPW [28] для LaFe$_{11.31}$Si$_{1.69}$ (M(Fe$_I$) = 2.07 $\mu_B$, M(Fe$_{II}$) = 2.42 $\mu_B$) и результатами нейтронографии [29] для LaFe$_{11.4}$Si$_{1.6}$ (M(Fe$_I$) = 1.59 $\mu_B$, M(Fe$_{II}$) = 2.12 $\mu_B$).

Типичные зависимости основных обменных интегралов от межатомного расстояния для LaFe$_{11.596-y}$Cr$_y$Si$_{1.404}$ показаны на Рис. 9. Как видно из рисунка, обменные интегралы достаточно быстро уменьшаются с увеличением межатомного расстояния и не превышают 1 мэВ уже на расстоянии 0.5 *a*. Наиболее значимыми являются обменные взаимодействия между атомами Fe$_{II}$ (~22 мэВ). Взаимодействие Fe$_{II}$-Fe$_I$ (между оболочкой икосаэдра и его центром) примерно в 3 раза меньше (~7 мэВ). Магнитные моменты атомов хрома направлены противоположно результирующему магнитному моменту элементарной ячейки. Величина обменного взаимодействия между ближайшими атомами Cr и Fe$_{II}$ составляет ~30 meV в LaFe$_{11.296}$Cr$_{0.3}$Si$_{1.404}$ и незначительно уменьшается при сжатии решетки или увеличении концентрации Cr. Обменное взаимодействие Cr-Cr также довольно велико (от -20 до -37 meV) и возрастает при увеличении концентрации хрома. Таким образом, легирование хромом в системе LaFe$_{11.596-y}$Cr$_y$Si$_{1.404}$ приводит к уменьшению общей намагниченности и способствует снижению $T_C$, что также наблюдалось экспериментально в [7].

В отличие от Cr, имеющего на два 3*d*-электрона меньше, чем у железа, Co имеет на один 3*d*-электрон больше и, соответственно больший ионный радиус, а значит добавление Co должно приводить к увеличению размеров элементарной ячейки и росту $T_C$, что наблюдалось экспериментально в [32]. Мы провели расчет зависимости полной энергии от размера элементарной ячейки для двух вариантов расположения атомов Co. В первом они считались случайно распределенными

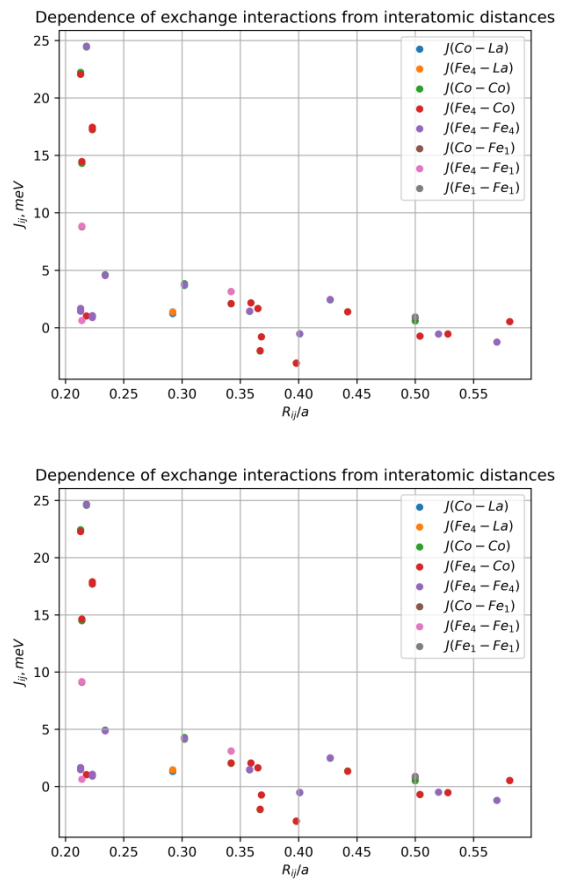

Рис.11. Зависимость величин интегралов межатомных взаимодействий от межатомного расстояния для La(Fe$_{0.352}$Co$_{0.648}$)$_1$(Fe$_{0.908}$Si$_{0.092}$)$_{12}$ при постоянной решетки a=6.033 Å (сверху) и a=6.059 Å (снизу).

по позициям $Fe_I$, а во втором предполагалось, что атомы Co случайным образом распределены по позициям $Fe_{II}$. Сравнение результатов расчета для $LaFe_{11.248}Co_{0.648}Si_{1.104}$ приведено на Рис. 10. Как видно из рисунка, Co энергетически выгодно занимать позиции типа $Fe_I$. При равновесном объеме в ферромагнитной фазе $LaFe_{11.248}Co_{0.648}Si_{1.104}$ магнитный момент элементарной ячейки, составляет 53.04 $\mu_B$, магнитный момент La -0.485 $\mu_B$, $Fe_I$ 2.0 $\mu_B$, 2.34 $\mu_B$ для $Fe_{II}$, - 0.196 $\mu_B$ для Si. Магнитный момент атомов Co сонаправлен с магнитным моментом элементарной ячейки и составляет 1.467 $\mu_B$.

Типичные зависимости основных обменных интегралов от межатомного расстояния для $LaFe_{11.248}Co_{0.648}Si_{1.104}$ показаны на Рис. 11. Как видно из рисунка, обменные интегралы достаточно быстро уменьшаются с увеличением межатомного расстояния и не превышают 1 мэВ уже на расстоянии 0.5 *a*. Наиболее значимыми являются обменные взаимодействия между атомами $Fe_{II}$ (~25 мэВ). Незначительно от него отстает обменное взаимодействие $Co_I$-$Fe_{II}$ (~22 мэВ). Взаимодействие $Fe_{II}$-$Fe_I$ (между оболочкой икосаэдра и его центром) примерно в 3 раза меньше (~8 мэВ). Обменное взаимодействие Co-Co также довольно велико (~22 мэВ), а изменение объема элементарной ячейки слабо влияет на межатомные обменные взаимодействия (Рис. 11).

## ЗАКЛЮЧЕНИЕ

Образцы сплавов $LaFe_{11.6}Si_{1.4}$ и $LaFe_{11.78}Mn_{0.41}Si_{1.32}H_{1.6}$ показали высокую повторяемость $\Delta Q$-эффекта при повторном включении магнитного поля, что критически важно для систем магнитного охлаждения. Образцы $LaFe_{11.78}Mn_{0.41}Si_{1.32}H_{1.6}$ демонстрирует высокие значения изотермического МКЭ с максимальным $\Delta Q$ = 3400 Дж/кг в магнитном поле $\mu_0 H$ = 1.8 Тл вблизи $T_C$ = 275 К, что в 2.5 раза превосходит аналогичные известные значения для чистого Gd при комнатной температуре.

Для исследования возможностей дальнейшего повышения функциональных свойств сплавов на основе $LaFe_{13-x}Si_x$ были проведены расчёты добавок Co и Cr. На основе проведенных расчетов показано, что добавление Cr приводит к уменьшению равновесного объема, т.е. к сжатию кристаллической решетки, что согласуется как с данными расчета методом FPLAPW, так и результатами нейтронографии. В отличие от Cr, имеющего на два 3*d*-электрона меньше, чем у железа, Co имеет на один 3*d*-электрон больше и, соответственно, больший ионный радиус, а значит добавление Co должно приводить к увеличению размеров элементарной ячейки. Таким образом, добавление Cr приводит к снижению, а Co – к росту $T_C$ в данных сплавах, что подтверждается экспериментально в работах [7] и [32], соответственно.

ПРИЛОЖЕНИЕ

# HIGH ISOTHERMAL MAGNETOCALORIC EFFECT IN La(Fe,Si)$_{13}$ BASED ALLOYS

A.P. Kamantsev[a,b], Yu. S. Koshkid'ko[a], O.E. Kovalev[b], N.Yu. Nyrkov[b], A.V. Golovchan[b], A.A. Amirov[c], A.M. Aliev[c]

[a] Kotelnikov Institute of Radioengineering and Electronics of RAS, Moscow, 125009 Russia

[b] FSBSI "Galkin Donetsk Institute for Physics and Technology", Donetsk, 283048 Russia

[c] Amirkhanov Institute of Physics, Dagestan Federal Research Center of RAS, Makhachkala, Republic of Dagestan, 367003 Russia

*e-mail: kaman4@gmail.com

This work investigates the magnetocaloric effect (MCE) in LaFe$_{11.6}$Si$_{1.4}$ and LaFe$_{11.78}$Mn$_{0.41}$Si$_{1.32}$H$_{1.6}$ alloys under adiabatic ΔT and isothermal ΔQ conditions in a magnetic field of $\mu_0 H = 1.8$ T. The studied samples exhibited high reproducibility of the ΔQ-effect upon cyclic magnetic field application, which is of critical importance for magnetic cooling systems. The LaFe$_{11.78}$Mn$_{0.41}$Si$_{1.32}$H$_{1.6}$ alloy demonstrate high values of the isothermal MCE, with a maximum ΔQ = 3400 J/kg in a magnetic field of $\mu_0 H = 1.8$ T near the Curie temperature ($T_C$) of 275 K. This value is 2.5 times higher than the well-known corresponding values for pure Gd at room temperature. Furthermore, the structural and magnetic properties of LaFe$_{13-x}$Si$_x$-based alloys with Cr and Co additions were investigated using density functional theory calculations. It was shown that the addition of Cr leads to a decrease in the equilibrium volume, i.e., to a compression of the crystal lattice, whereas Co addition causes its expansion. These changes are expected to increase or decrease the $T_C$, respectively.

**Keywords:** magnetocaloric effect, La-Fe-Si alloys, high magnetic fields, isothermal heat

# ПОДПИСИ К РИСУНКАМ

**Рис. 1.** Рис. 1. Температурные зависимости адиабатического изменения температуры $\Delta T$ для $LaFe_{11.6}Si_{1.4}$, отожженного при 1323 K в течение 7 суток, данные из [12].

**Рис. 2.** Рис. 2. Температурные зависимости адиабатического изменения температуры $\Delta T$ для $LaFe_{11.78}Mn_{0.41}Si_{1.32}H_{1.6}$ при нагреве и охлаждении.

**Рис. 3.** Рис. 3. Температурные зависимости изотермического тепловыделения $\Delta Q$ образца $LaFe_{11.6}Si_{1.4}$ при охлаждении и нагреве. Измерения для каждой температуры проводились дважды (первое и второе включение поля).

**Рис. 4.** Рис. 4. Температурные зависимости изотермического тепловыделения $\Delta Q$ образца $LaFe_{11.78}Mn_{0.41}Si_{1.32}H_{1.6}$ при охлаждении и нагреве.

**Рис. 5.** Температурные зависимости изотермического изменения энтропии $\Delta S$ для $LaFe_{11.6}Si_{1.4}$ и $LaFe_{11.78}Mn_{0.41}Si_{1.32}H_{1.6}$ при охлаждении и нагреве в магнитном поле 1.8 Тл.

**Рис. 6.** Рис. 6. Магнитополевые зависимости изотермического изменения тепла $\Delta Q$ для $LaFe_{11.6}Si_{1.4}$ при 1-ом включении поля в режимах (a) нагрева и (b) охлаждения.

**Рис. 7.** Кристаллическая структура $LaFe_{13-x}Si_x$ и основные межатомные обменные интегралы, где $J_1$ – обменный интеграл между $Fe_{II}$-$Fe_{II}$ из соседних икосаэдров, $J_2$ – обменный интеграл между $Fe_I$-$Fe_{II}$ (между оболочкой икосаэдра и его центром), $J_3$ – это обменный интеграл между $Fe_{II}$-$Fe_{II}$ из соседних икосаэдров, $J_4$ – обменный интеграл между $Fe_{II}$-$Fe_{II}$.

**Рис. 8.** Влияние Cr на магнитный момент элементарной ячейки и зависимость полной энергии от постоянной кристаллической решетки в $LaFe_{11.596-y}Cr_ySi_{1.404}$.

**Рис. 9.** Зависимость величин интегралов межатомных взаимодействий от межатомного расстояния для $LaFe_{11.596}Si_{1.404}$ (сверху), $LaFe_{11.296}Cr_{0.3}Si_{1.404}$ (в середине) и $LaFe_{10.996}Cr_{0.6}Si_{1.404}$ (снизу). Во всех случаях постоянная кристаллической решетки $a = 6.06$ Å.

**Рис. 10.** Влияние распределения Co по позициям железа на зависимость магнитного момента элементарной ячейки и полной энергии от постоянной кристаллической решетки.

**Рис. 11.** Зависимость величин интегралов межатомных взаимодействий от межатомного расстояния для $La(Fe_{0.352}Co_{0.648})_1(Fe_{0.908}Si_{0.092})_{12}$ при постоянной решетки a=6.033 Å (сверху) и a=6.059 Å (снизу).